\documentclass[conference, letterpaper]{IEEEtran}
\IEEEoverridecommandlockouts

\usepackage[english]{babel}
\usepackage[caption=false]{subfig}
\usepackage{floatrow}
\usepackage[latin1]{inputenc}
\usepackage{bbm}
\usepackage{url}
\usepackage{amsmath,pifont,amssymb,amsfonts}
\usepackage{graphicx}
\usepackage{color}
\usepackage[table]{xcolor}
\usepackage{framed}
\usepackage{gensymb}
\usepackage{longtable}
\usepackage{array}
\usepackage{caption}
\usepackage{psfrag}
\usepackage{algorithm}
\usepackage{algpseudocode}
\usepackage{cite}
\usepackage{multirow}
\usepackage{balance}

\graphicspath{{figs/}}

\begin{document}
\title{DeepMAL - Deep Learning Models for Malware Traffic Detection and Classification}
\author{\IEEEauthorblockN{Gonzalo Mar\'in\IEEEauthorrefmark{1}\IEEEauthorrefmark{2}\IEEEauthorrefmark{3}, Pedro Casas\IEEEauthorrefmark{1}, Germ\'an Capdehourat\IEEEauthorrefmark{2}\IEEEauthorrefmark{4}}
\IEEEauthorblockA{\IEEEauthorrefmark{1}AIT Austrian Institute of Technology, Vienna, Austria\\
\IEEEauthorrefmark{2}IIE--FING, Universidad de la Rep\'ublica, Montevideo, Uruguay\\
\IEEEauthorrefmark{3}Tryo labs, Montevideo, Uruguay\\
\IEEEauthorrefmark{4}Plan Ceibal, Montevideo, Uruguay\\
gonzalo.marin@fing.edu.uy, pedro.casas@ait.ac.at, gcapdehourat@ceibal.edu.uy}
}

\maketitle
\begin{abstract}
Robust network security systems are essential to prevent and mitigate the harming effects of the ever-growing occurrence of network attacks. In recent years, machine learning-based systems have gain popularity for network security applications, usually considering the application of \emph{shallow} models, which rely on the careful engineering of expert, handcrafted input features. The main limitation of this approach is that handcrafted features can fail to perform well under different scenarios and types of attacks. Deep Learning (DL) models can solve this limitation using their ability to learn feature representations from \emph{raw}, non-processed data. In this paper we explore the power of DL models on the specific problem of detection and classification of malware network traffic. As a major advantage with respect to the state of the art, we consider raw measurements coming directly from the stream of monitored bytes as input to the proposed models, and evaluate different raw-traffic feature representations, including packet and flow-level ones. We introduce DeepMAL, a DL model which is able to capture the underlying statistics of malicious traffic, without any sort of expert handcrafted features. Using publicly available traffic traces containing different families of malware traffic, we show that DeepMAL can detect and classify malware flows with high accuracy, outperforming traditional, shallow-like models.
\end{abstract}
\IEEEpeerreviewmaketitle

\section{Introduction}

The popularity of Artificial Intelligence (AI) - and of Machine Learning (ML) as an approach to AI, has dramatically increased in the last few years due to its outstanding performance in various domains, notably in image, audio, and natural language processing, and more recently in gaming. A major breakthrough in AI/ML has been the successful application of Deep Learning (DL) models and techniques to traditional ML tasks. One of the most powerful characteristics of DL models is their ability to learn feature representations from input raw or basic, non-processed data. For example, a Convolutional Neural Network (CNN) trained for image classification can learn to recognize edges and more complex structures along the sequence of neural layers, using as input only the image pixel values \cite{DBLP:journals/corr/SimonyanVZ13}. The key behind DL is the so-called \emph{representation learning} paradigm \cite{Bengio2013}, which offers a set of methods allowing a ML algorithm to automatically discover the best data representations or features from raw data inputs. DL models are basically representation learning methods with multiple levels of abstraction, obtained by composing simple but non-linear consecutive transformation layers, each of them providing a more abstract representation of the data.

Despite the success of DL models, shallow ML models are usually applied when it comes to the analysis of network traffic measurements. Shallow ML models rely on the definition of highly specialized, expert-handcrafted features to achieve proper results, and this is actually critical for shallow ML. There are different problems when addressing network traffic measurements tasks using shallow ML approaches. First, the lack of a consensual labeled \textit{raw} traffic, full-packet capture dataset to train ML models (e.g. due to privacy policies in the data); second, the lack of a consensual set of input features to tackle specific targets, such as network security, anomaly detection, traffic classification, etc.; third, the continuous changing in network measurements statistics that may cause static handcrafted features to fail. To improve these limitations, we explore the end-to-end application of DL models to complement traditional approaches for network measurement analysis, using different representations of the input data. We introduce DeepMAL, a DL model capable to detect and classify malware network traffic using raw, bytestream-based data as input. We describe and evaluate different DL architectures and different input representations, showing an outstanding detection performance through the analysis of raw bytestream packet data.

The rest of the paper is organized as follows: in Section \ref{sec:state} we present a brief state-of-the-art on DL models applied to the analysis of network traffic measurements; in Section \ref{sec:dl} we present and provide details on the different DL approaches and architectures behind DeepMAL; in Section \ref{sec:results} we discuss the detection performance achieved by different versions of DeepMAL, comparing them with traditional, shallow-like based models, using domain expert knowledge to craft the input features; in Section \ref{sec:multi} we introduce a variation of the detection problem using a multi-class, malware classification approach, and present concluding remarks in Section \ref{sec:conclu}.

\section{State-of-the-art}\label{sec:state}

The application of shallow, ML models to general network measurement problems is largely extended in the literature. There are a couple of extensive surveys and papers on network measurement problems such as network anomaly detection \cite{survey_ahmed} \cite{survey_adnet} - including ML-based approaches \cite{ml_ad}, ML for network traffic classification \cite{MLsurvey} and network security \cite{mlsec_sur}. A recent survey on ML for networking is presented in \cite{Boutaba2018}, discussing different applications and associated challenges.

DL approaches have started to be used recently with promising performance results, mainly associated to traffic classification tasks. In 2015, Z. Wang \textit{et al.} \cite{WangTheAO} presented a Deep Neural Network for feature learning using feed-forward networks and Stacked Auto-Encoders (SAE) to perform network protocol recognition over a dataset made up of TCP flows from an internal network. In 2017 there have been some works over the subject. W. Wang \textit{et al.} presented two models based upon a 2D-CNN \cite{7899588} and 1D-CNN \cite{8004872}, in which the authors transform network flows and sessions to images -- as done by \cite{r6}, to work as an input for the CNN models, using either the information of all the layers, or only from the application layer. A similar image-based approach to traffic classification was recently introduced in \cite{flowpic}. Lotfollahi \textit{et al.} \cite{DBLP:journals/corr/abs-1709-02656} presented an approach for encrypted traffic classification using SAE and 1D-CNN at the packet level. M. Lopez-Martin \textit{et al.} \cite{8026581} presented different DL architectures based on CNN and LSTM networks to perform traffic classification using self-collected network traces. Additional papers based on DL architectures for traffic classification cover the classification of mobile networks encrypted traffic \cite{r1,r3} and the classification of general network traffic \cite{r6,r10}.

When it comes to intrusion and anomaly detection, there are also many recent papers applying DL architectures. Radford \textit{et al.} \cite{DBLP:journals/corr/abs-1803-10769} presented an anomaly detection model using a LSTM network from network traffic logs for cyber-security. Papers such as \cite{r4,r7} rely on the application of CNNs to capture the underlying spatial correlations of the analyzed data for intrusion detection, using already pre-processed input features -- e.g., \cite{r4,r7} use the well-known KDD features and dataset. In \cite{r8}, J. Cui \textit{et al.} proposed a word embedding-based approach to pre-analyze the raw traffic data for network intrusion detection. H. Huang \textit{et al.} apply a multi-task learning approach to detect intrusions and anomalies, using again CNNs. W. Wang \textit{et al.} \cite{r2} propose a hierarchical learning approach to construct spatio-temporal features, combining spatial feature learning through CNNs and temporal feature learning through recursive networks. Finally, \cite{marin_rawpower_2018, marin_wain_2018} introduces \textit{RawPower}, a DL architecture for network anomaly and intrusion detection, which combines spatio and temporal feature learning for detection. Most of these proposals use DL models after some sort of preprocessing of the data, or based on the definition of handcrafted features to generate DL model inputs. 

The main contribution of our approach is that of using completely expert-knowledge-independent inputs for both the prediction and classification tasks - just the raw bytestream, opening the door to a broad set of potential applications of DL in networking problems. Additional contributions include: (i) the benchmarking of DeepMAL against shallow-like learning models commonly used in the literature; (ii) addressing not only the problem of binary malware detection, but also the classification of the specific class of malware detected; (iii) the evaluation of DeepMAL on multiple different network traffic datasets, publicly available from the well-known Stratosphere IPS Project of the CTU University of Prague in Czech Republic \cite{Garcia:2014:ECB:2664689.2665897}.

\begin{figure}[t!]
\centering
    \subfloat[\label{fig:packetrep} Packet representation for the input data. The shape of the input data is $(N,n)$: $N$ is the number of packets and $n$ the number of bytes.]{\includegraphics[trim={0.1cm 0.1cm 0.1cm 0.1cm},clip,width=0.475\textwidth]{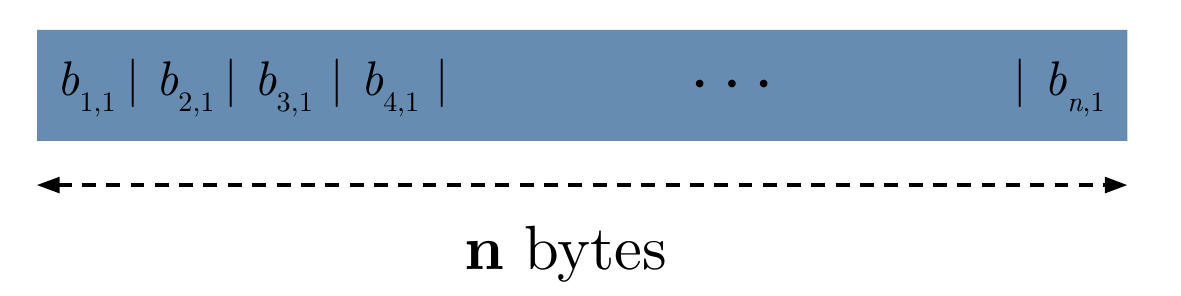}}
    \quad
    \subfloat[\label{fig:flowrep} Flow representation for the input data. A tensor of size $(N,m,n)$ where $N$ represents the number of flows, $m$ the number of packets, and $n$ the number of bytes.]{\includegraphics[trim={0.1cm 0.1cm 0.1cm 0.1cm},clip,width=0.475\textwidth]{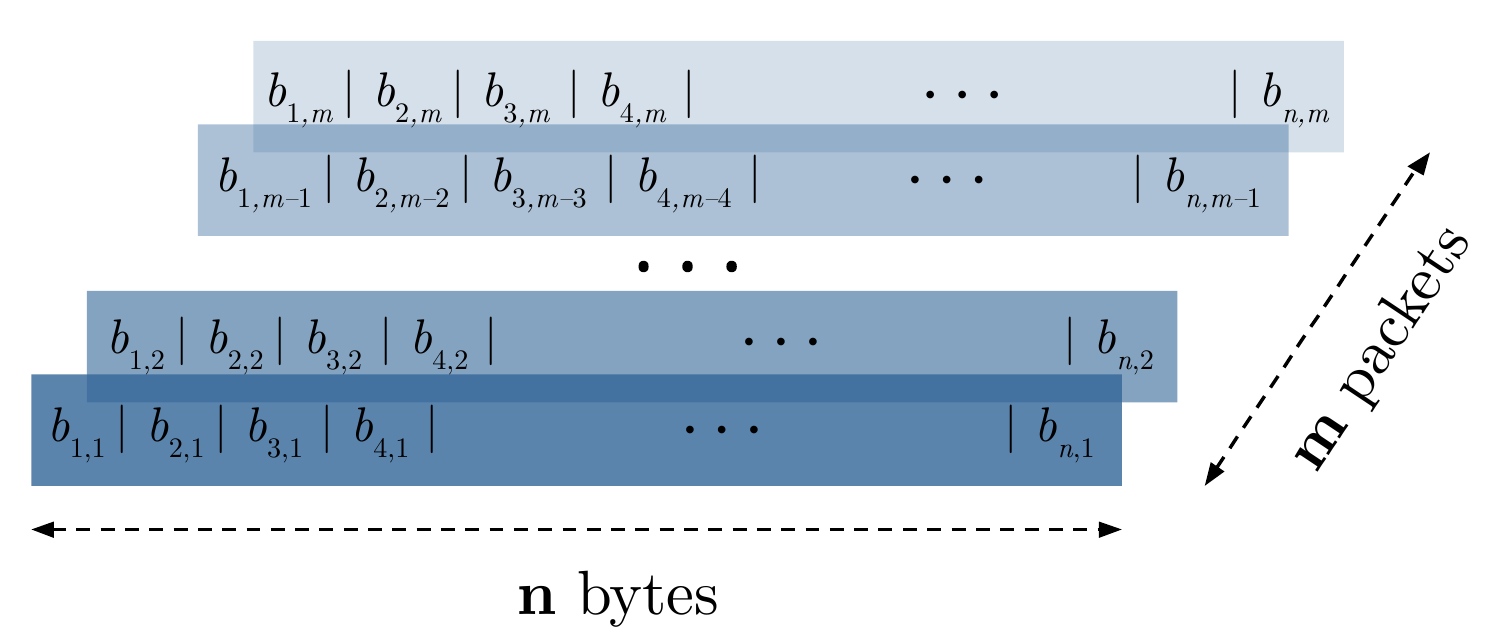}}
    \caption{Different input representation for DeepMAL.}\label{fig:inputrep}
\end{figure}

\begin{figure*}[t!]
\centering
\includegraphics[width=\textwidth]{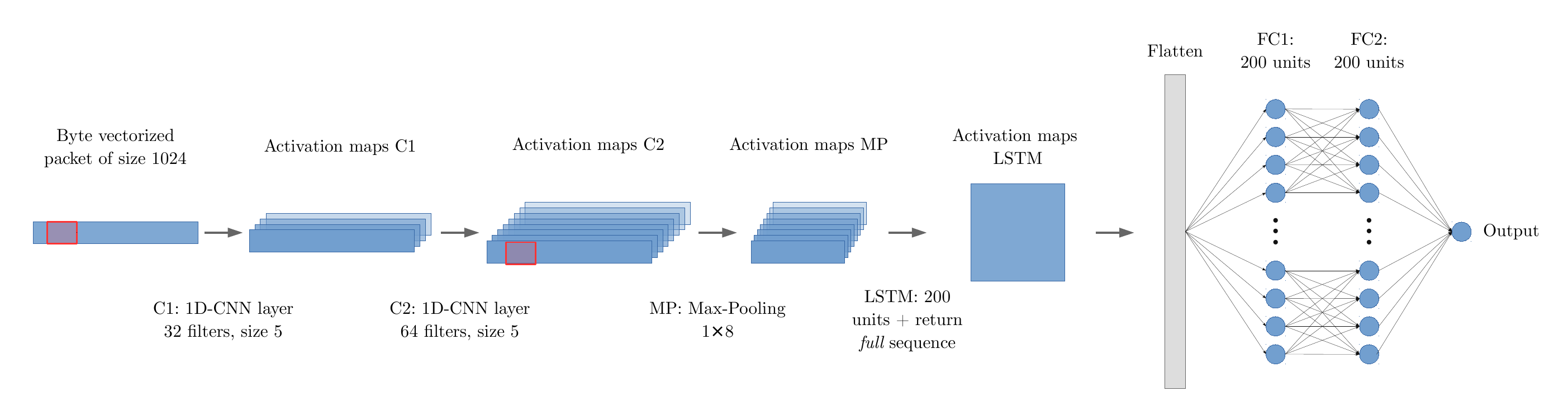}
\caption{DL architecture for \textit{Raw Packets} representation.}\label{fig:pktarch}
\end{figure*}

\begin{figure}[t!]
\centering
\includegraphics[width=\textwidth]{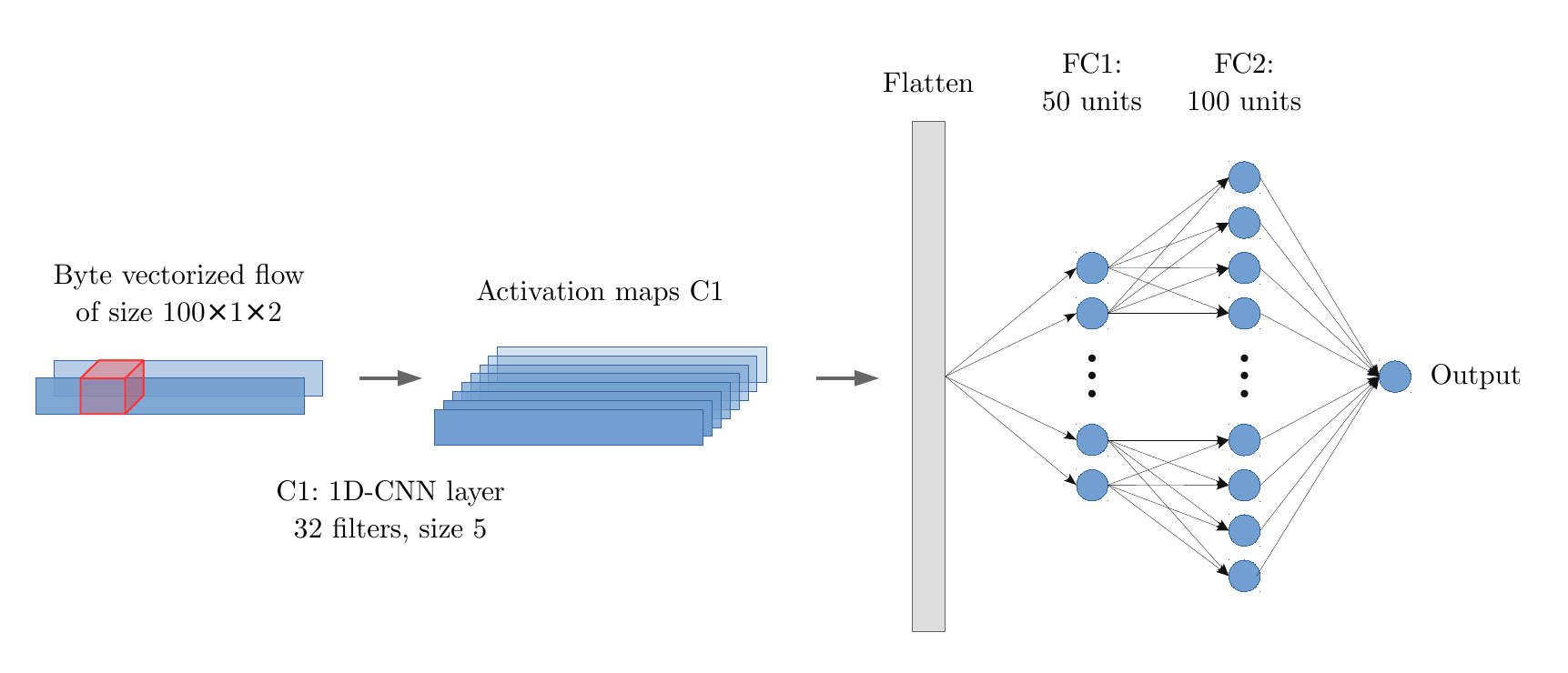}
\caption{DL architecture for \textit{Raw Flows} representation.}\label{fig:flowsarch}
\end{figure}

\section{DL for Malware Detection}\label{sec:dl}

Our goal is to train a DL model with the stream of incoming bytes without requiring any preprocessing step or domain expert intervention, to make the approach generic and flexible. As we said before, we particularly focus on the problem of malware traffic detection and classification, using raw, bytestream-based data as input. The input representation of the data, as well as the network architecture, are both key elements to consider when building a DL model. Since we want to evaluate the feature representation power of the model from non-processed data, we consider two types of raw representations: packets and flows. In both cases we consider decimal normalized representation of every byte of every packet as a different feature. Since different packets can have different sizes, we set a fixed threshold $n$ to trim each incoming packet to the first $n$ bytes, after removing potentially biasing byte information such as MAC and IP addresses. All packets with size larger than $n$ bytes are trimmed, and packets with smaller size are zero-padded. Next we describe basic concepts behind DeepMAL's architecture, and present the two raw representations based on packets and flows, along with the corresponding network architectures.

\subsection{DL Architecture Basics}\label{sec:architecture}

Finding the right DL architecture for a specific classification task is challenging. In general, one would like to capture different spatial and temporal correlations and phenomena hidden within the specific input data. To capture both types of correlations and learn spatio-temporal features, the proposed DL architectures consider both convolutional and recurrent neural networks as core layers. The convolutional layer(s) are included to build the feature representation of the spatial data inside packets and flows. The recurrent layers are used together with the convolutional ones to improve detection and classification performance, allowing the model to keep track of temporal information. We additionally consider fully-connected neural layers to deal with the different combinations of features, to arrive to the final decision -- i.e., classification rule.

Besides these core layers, we also include what we refer-to as \emph{helper layers}, which goal is to reduce the generalization error and to improve the learning process, reducing over-fitting issues. As helper layers, we consider batch normalization and dropout layers. Batch normalization is used to normalize layer inputs for each mini-batch during training; as a result, higher learning rates can be used, the resulting model is less sensitive to initialization, and it additionally contributes to the regularization of the model. The dropout layers are also used to regularize the model, by doing the so-called \emph{model averaging}; dropout consists of randomly dropping units -- along with their connections, from the neural network during training. In a nutshell, dropout is a very efficient way to perform model averaging, as it is similar to train a huge number of different networks and
average the results.

Next, we use different combinations of these core and helper layers to build the proposed architectures. In all cases, we tried different combinations and network sizes, and keep those combinations that provided the best empirical results. A deeper and more structured analysis of the interplays among these different layers is part of our ongoing work.

\subsection{Input Representations}\label{sec:repr}

In the packet approach, we consider each packet as a different instance, while in the flow approach we consider a group of packets -- that make up the flow -- as an input for the network. Both representations are depicted in Fig.~\ref{fig:inputrep}. For the \textit{Raw Packet} representation, we have to choose the number of bytes from the packet to consider ($n$), while in the flow representation we also have to set the number of packets per flow to consider ($m$). This is because, naturally, different packets and flows can have different sizes.

Since both malware and normal captures are gathered under controlled conditions, there is some bias in the IP and transport protocol headers that are not representative of \textit{in the wild} traffic. This is the case, for example, of fixed values for IP addresses and ports and even some of the transport protocol flags. For this reason, we take the \textit{payload} of every packet as the key information to analyze and to build the dataset. Afterwards, we set a fixed threshold for the parameter $n$ to trim each incoming packet to the first $n$ bytes of payload. All packets with size larger than $n$ bytes are trimmed, and packets with smaller size are zero-padded at the end. For the number of packets per flow, we fix a number $m$ and take the first $m$ packets of the flow, discarding the rest.

\subsection{DeepMAL using Raw Packets}

The architecture of DeepMAL for the \textit{Raw Packets} input representation is shown in Fig.~\ref{fig:pktarch}. It consists of two 1D-CNN convolutional layers of 32 and 64 filters of size 5, respectively; a max-pooling layer of size 8 to down-sample the input; a well-known LSTM as recursive layer, consisting of 200 units, returning the outputs of each cell (``return sequences'' mode on); and finally, two fully-connected (FC) layers of 200 units each. A binary cross-entropy is used as the loss function. Spatial and normal batch normalization layers are added after each 1D-CNN and FC layers to ease the training process. Dropout layers are also used to add regularization to the model.

\subsection{DeepMAL using Raw Flows}

When deciding on the network architecture for the \textit{Raw Flows} approach, we note that the number of instances to deal with when operating at the flow level is by far much smaller than in the case of packet-based inputs; as a consequence, the capacity of the model does not have to be as high as in the \textit{Raw Packets} case. For this reason, we only include convolutional layers in this architecture. The final architecture in this case consists of one single 1D-CNN layer of 32 filters of size 5 and two fully-connected layers of 50 and 100 units each. Also, binary cross-entropy is used as the loss function. The architecture is shown in Fig.~\ref{fig:flowsarch}.

\section{Experimental Evaluations}\label{sec:results}

We evaluate the different proposed DeepMAL architectures and input representations using real network measurements, publicly available though the \textit{Stratosphere IPS Project} of the CTU University of Prague in Czech Republic \cite{Garcia:2014:ECB:2664689.2665897}. In this section we focus exclusively on the problem of malware detection, posing it as a binary classification task: either normal instance or malware. To show the main advantages of DeepMAL, we pose ourselves three evaluation questions: (i) is it possible to achieve high detection accuracy with low false alarm rates using the raw-input, DL-based models?; (ii) are the proposed DL-based models better than the commonly used shallow models for malware detection, when feeding them all with raw inputs (e.g., byte-streams)?; and (iii) how good are the raw-input, DL-based models as compared to a traditional ML-based approach for malware detection, where shallow models take as input specific hand-crafted features based on domain expert knowledge?.  

\begin{figure}[t!]
\centering
$\begin{array}{c}
\includegraphics[width=0.75\columnwidth]{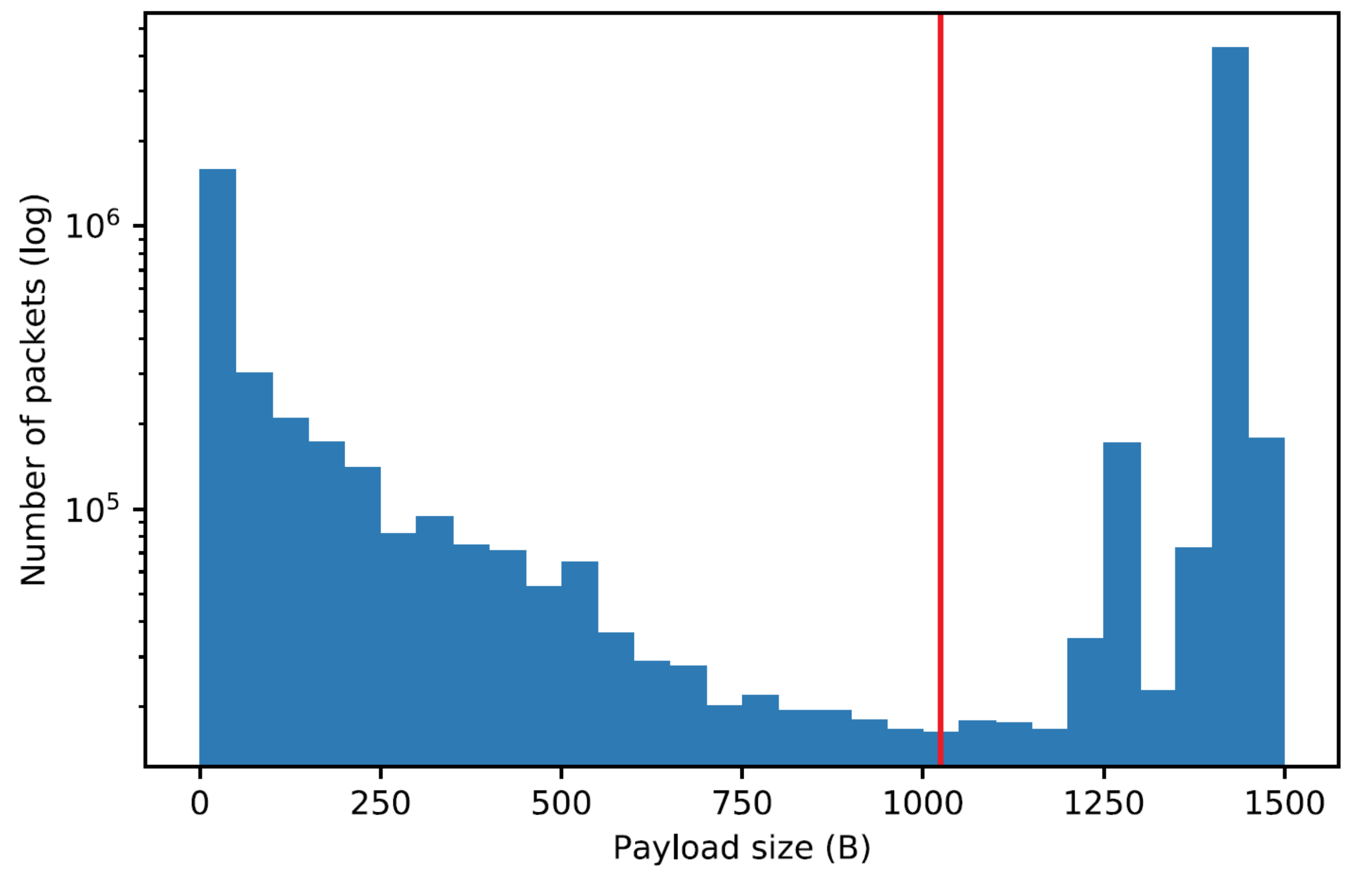}\\
\text{(a) Normal traffic.}\\
\includegraphics[width=0.75\columnwidth]{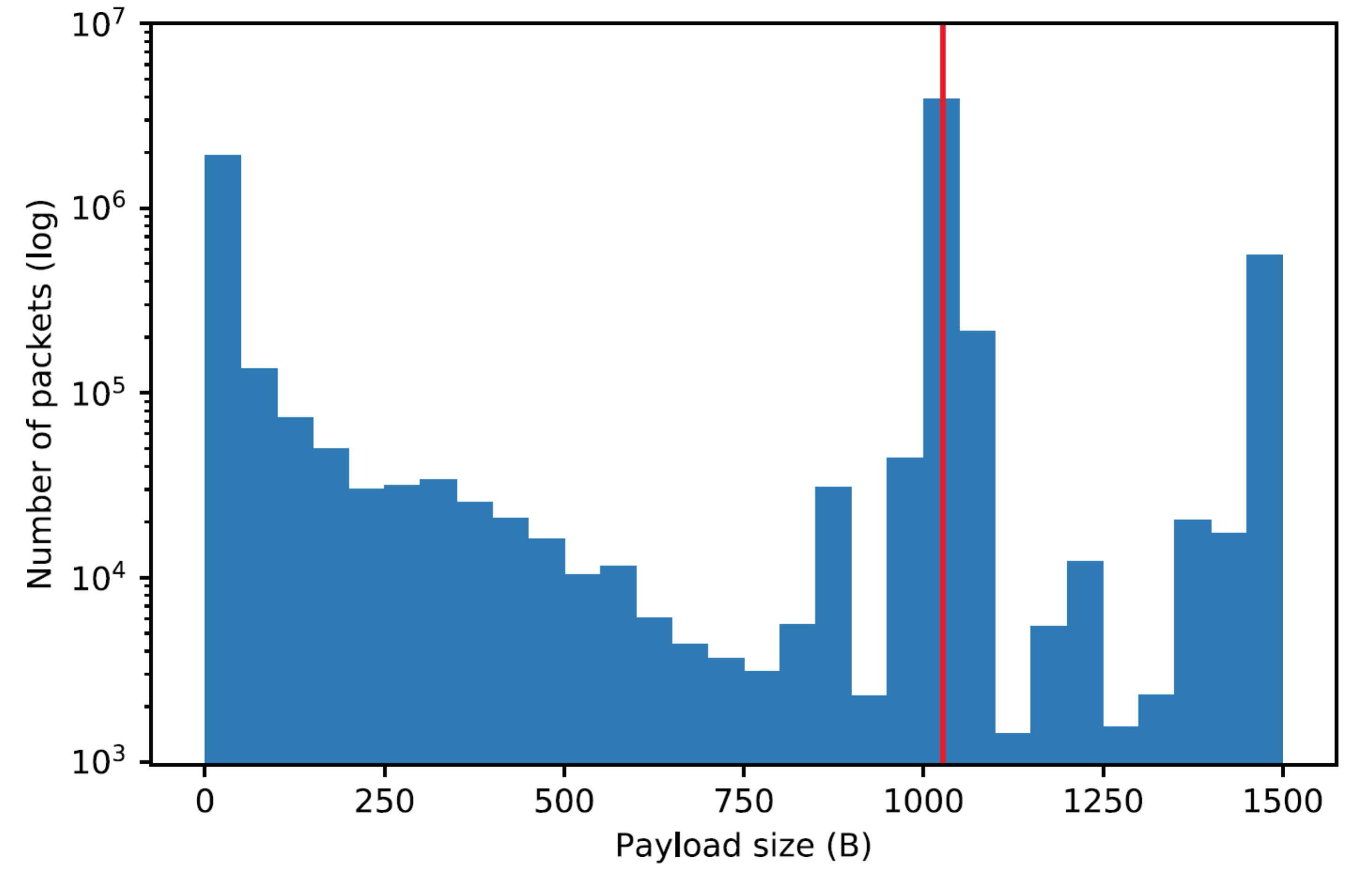}\\
\text{(b) Malware.}
\end{array}$
\caption{Payload size for normal and malware packets.}
\label{fig:pkt_bytes}
\end{figure}

\begin{figure}[t!]
\centering
\includegraphics[width=0.9\columnwidth]{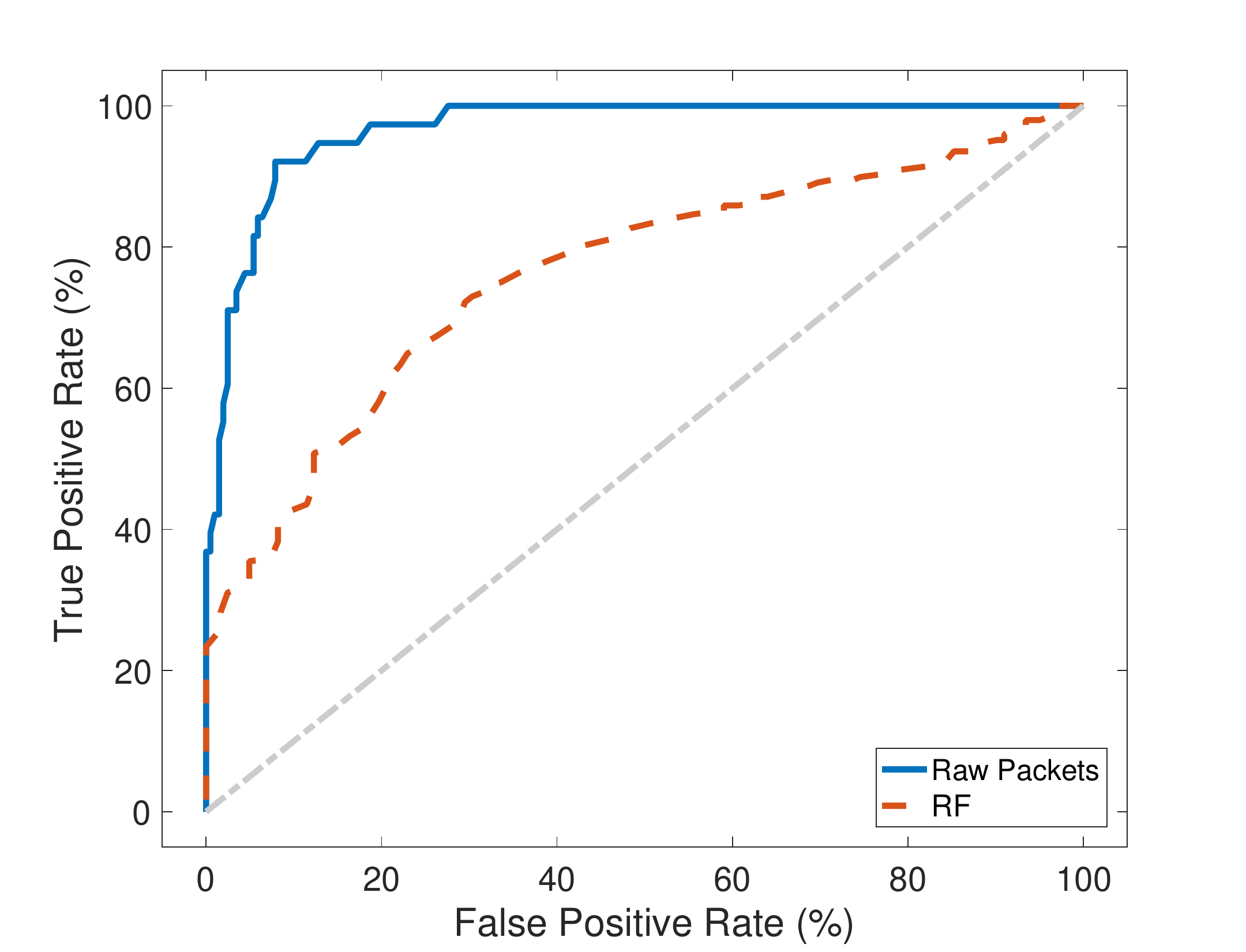}
\caption{Malware detection performance. Raw packet inputs, DeepMAL vs. RF model.}\label{roc_results}
\end{figure}

\begin{figure*}[t!]
\centering
    \subfloat[\label{fig:pktlossiter} \textit{Raw Packets} representation.]{\includegraphics[trim={0.1cm 0.1cm 0.1cm 
    0.1cm},clip,width=.465\textwidth]{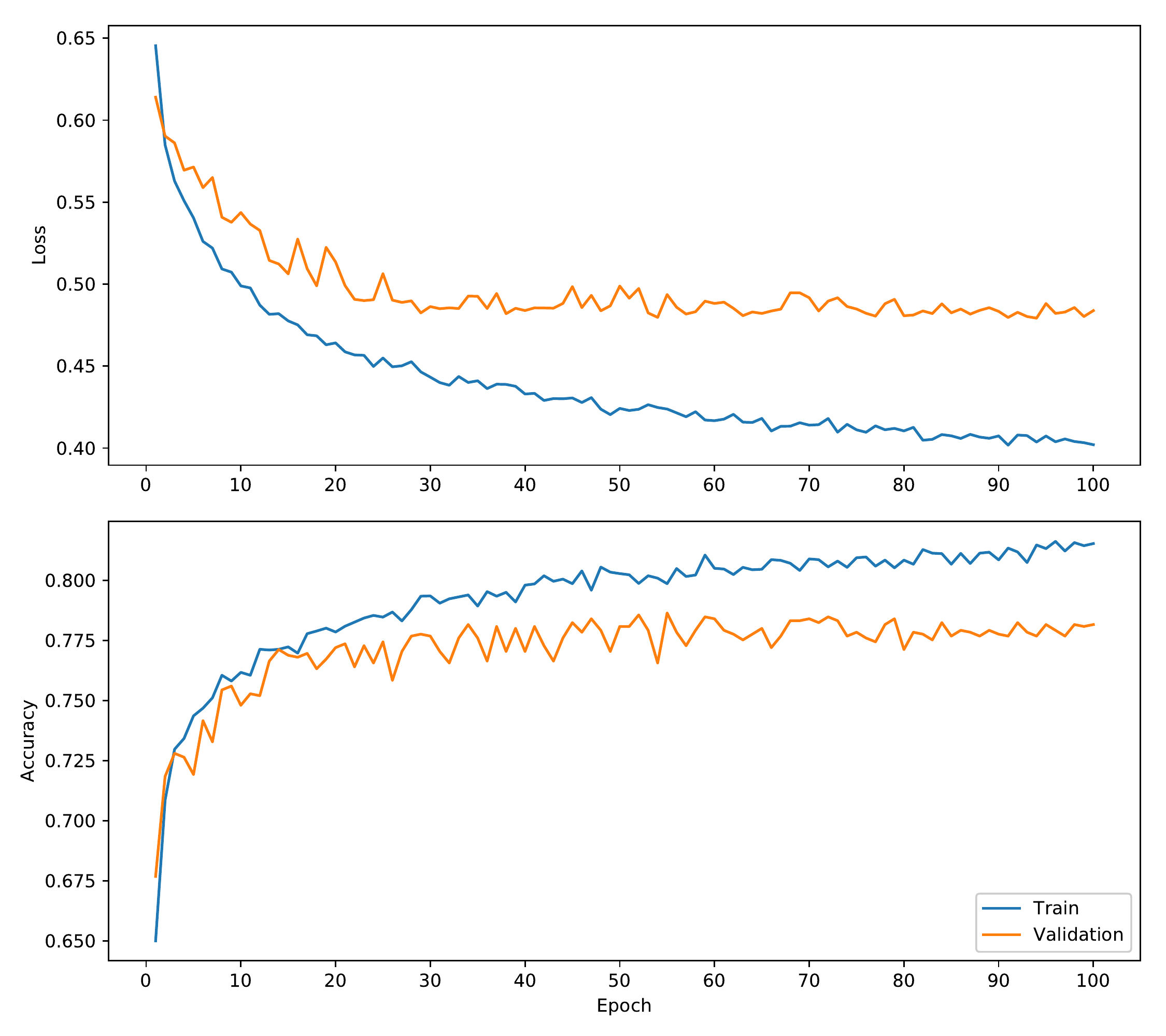}}
    \quad
    \subfloat[\label{fig:flowlossacc} \textit{Raw Flows} representations.]{\includegraphics[trim={0.1cm 0.1cm 0.1cm
    0.1cm},clip,width=.465\textwidth]{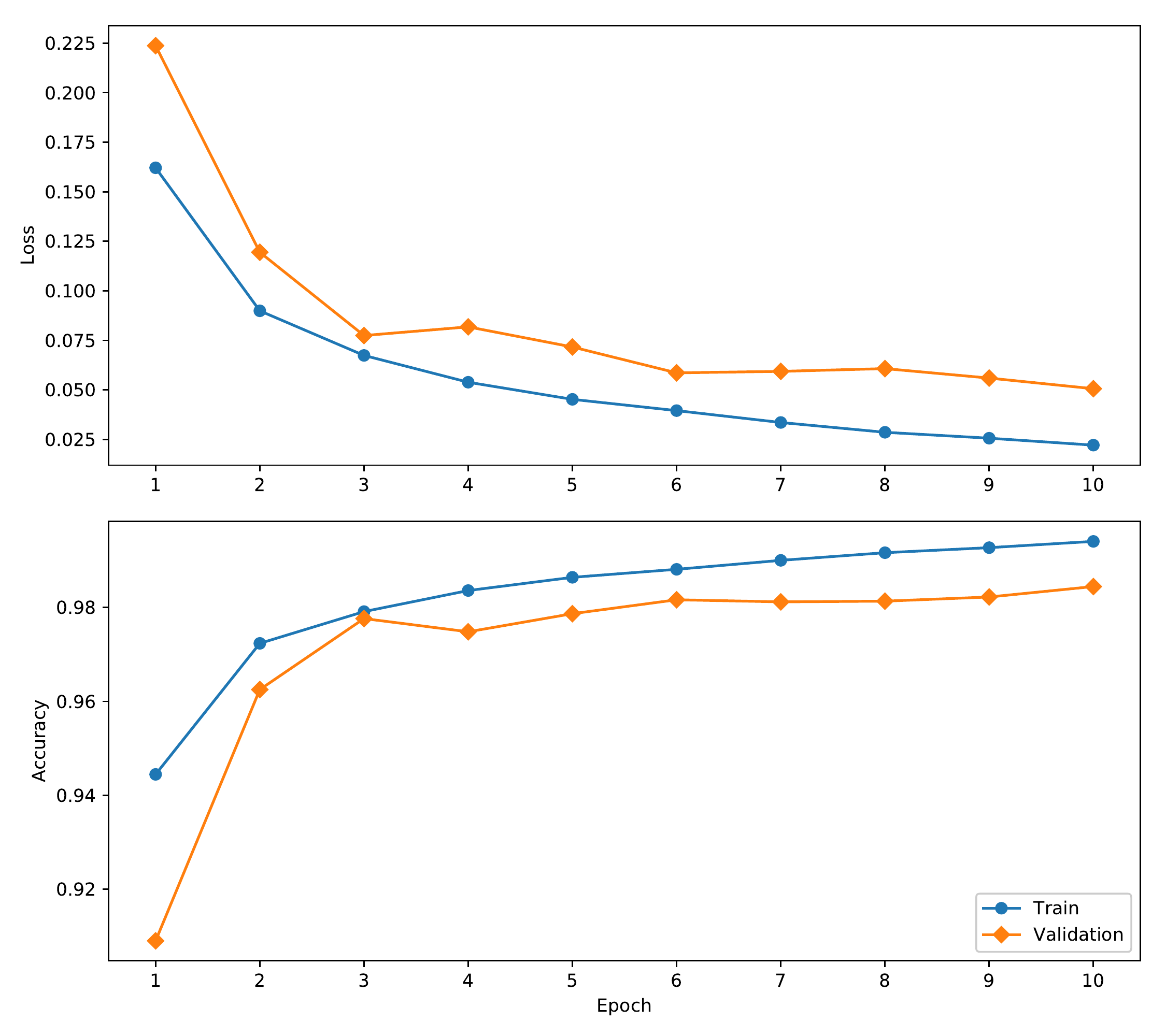}}
    \caption{Learning performance (loss and accuracy evolution after each epoch) for \textit{Raw Flows} and \textit{Raw Packets} approaches.}\label{fig:learning}
\end{figure*}

\begin{figure*}[t!]
\centering
    \subfloat[\label{fig:rocpackets}\textit{Raw Packets} representation.]{\includegraphics[width=.465\columnwidth]{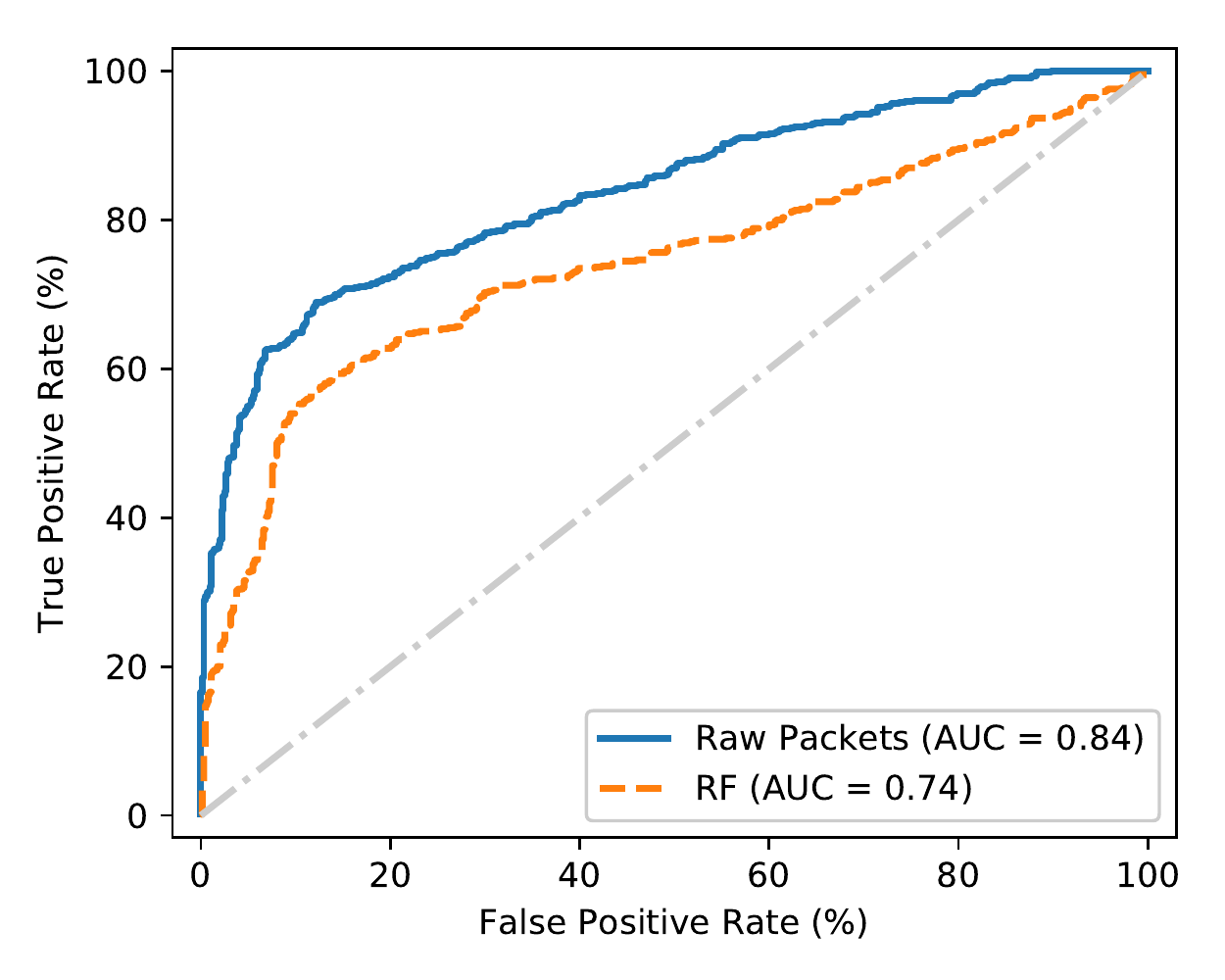}}
    \quad
    \subfloat[\label{fig:rocflows}\textit{Raw Flows} representation.]{\includegraphics[width=.465\columnwidth]{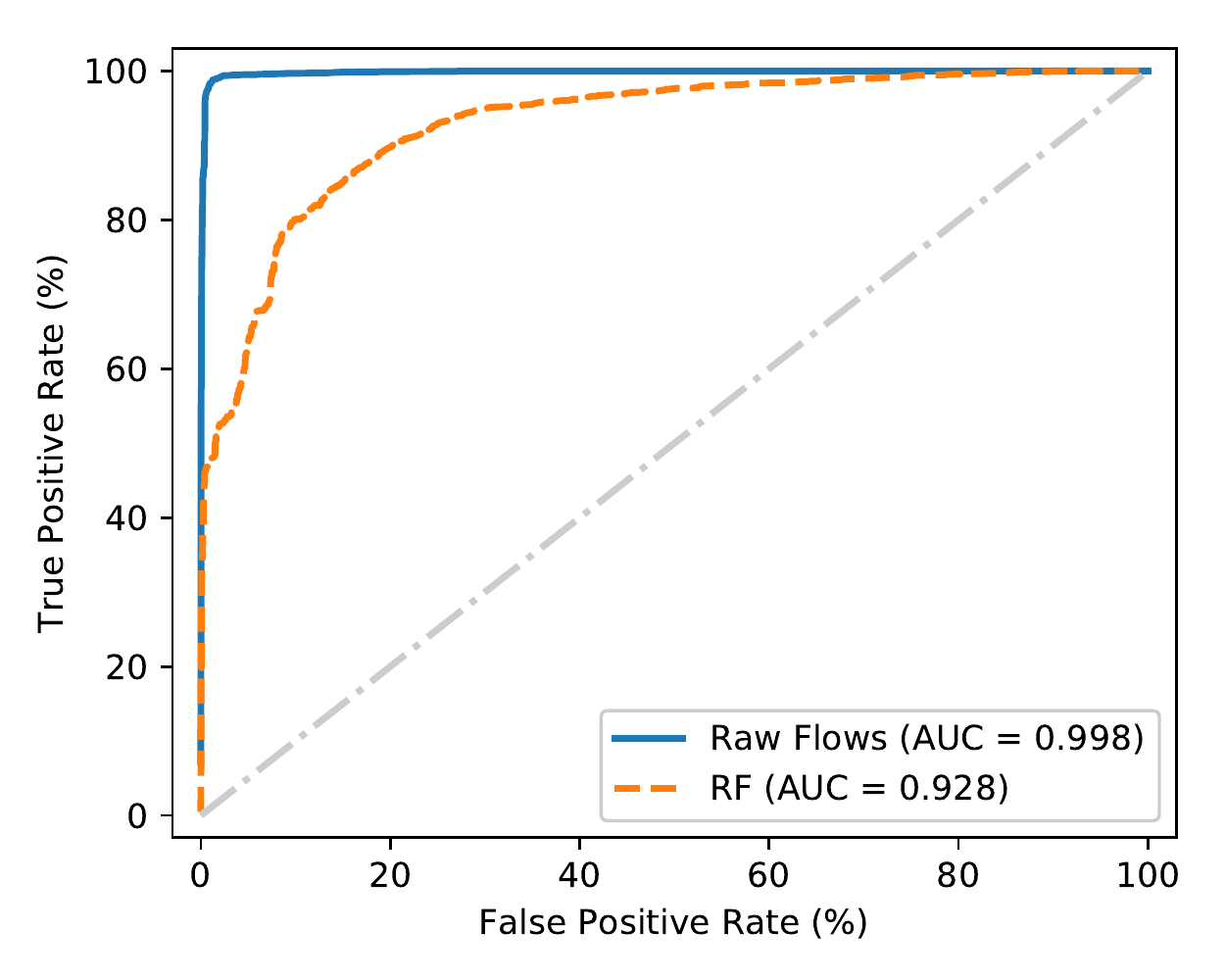}}
    \caption{Malware detection performance using \textit{Raw Packets} and \textit{Raw Flows} representations.}\label{fig:rocs}
\end{figure*}

\begin{figure}[t!]
\centering
\includegraphics[width=0.925\columnwidth]{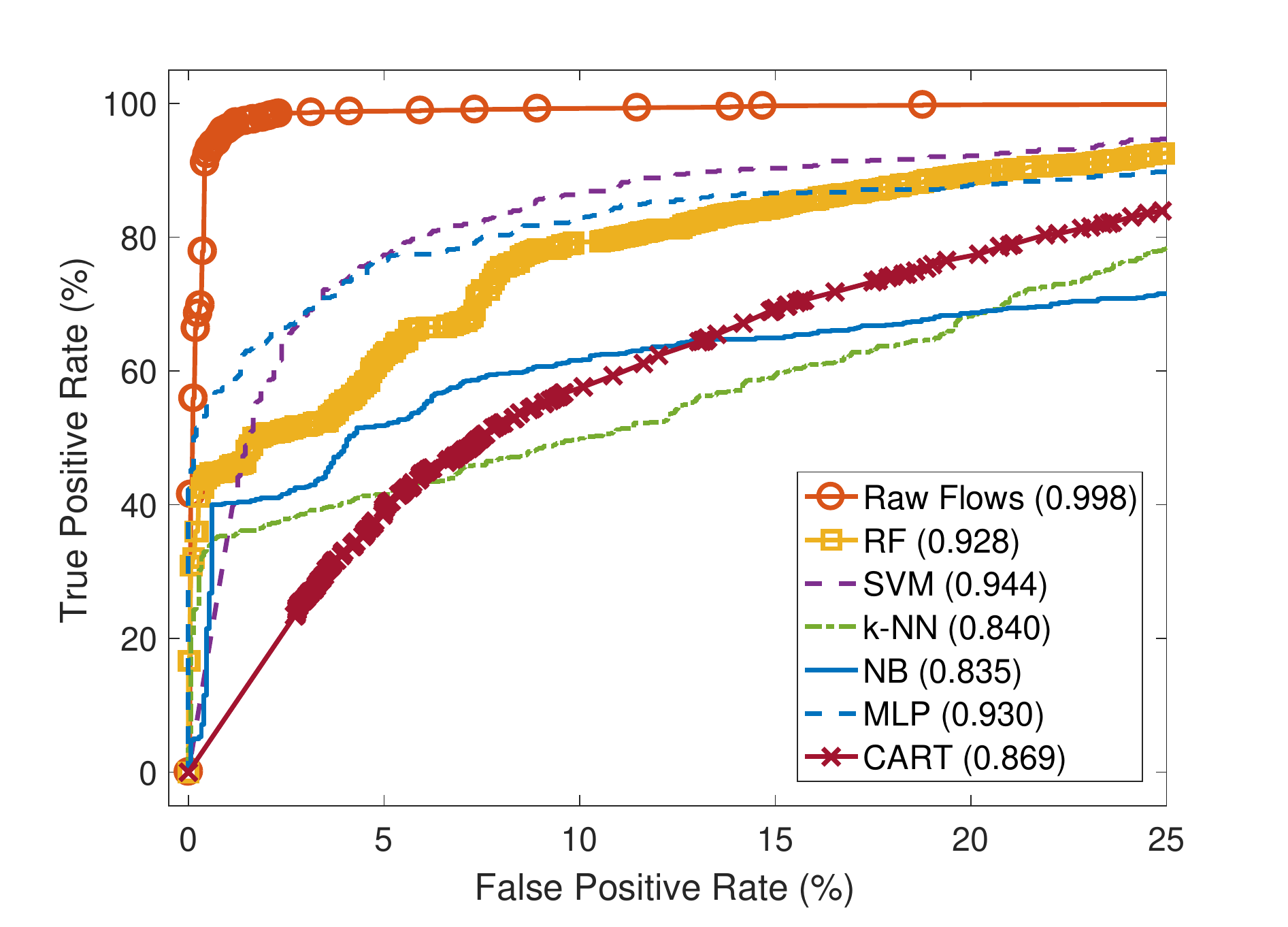}
\caption{\textit{Raw Flows} vs. different shallow models.}
\label{fig:rocs_all_cnn}
\end{figure}

\subsection{DeepMAL vs. Shallow Models with Raw Inputs}

We begin with a simple evaluation scenario, detecting malware at the packet level. We take a first dataset previously used in the literature \cite{7899588}, referred to as the \texttt{USTC\-TFC2016} dataset. This dataset consists of two groups of labeled \texttt{pcap} files from malware and benign applications. The malware files contain 10 types of malware traffic from public websites, collected from a real network environment by researchers of the Czech Technical University in Prague (CTU) \cite{Garcia:2014:ECB:2664689.2665897}. The benign traffic contains 10 types of normal traffic which were collected using Ixia BreakingPoint (\url{https://www.ixiacom.com}), a network traffic simulation platform. We take every packet from each labeled \texttt{pcap} as part of the dataset, using decimal normalized representation of every byte as a different feature. 

To set the number of bytes kept from each packets, we resort to a simple, data-driven analysis of the dataset. Fig. \ref{fig:pkt_bytes} depicts the distribution of the packets payload size for (a) all the normal or benign traffic samples, and (b) all the malware packets. As clearly observed, both distributions are rather similar for payload sizes below 750 bytes and above 1250 bytes, but markedly different between these two boundaries. We therefore decided to set $n = 1024$ bytes, which proved to be a reasonably good threshold to capture the difference between benign and malware activity. Each packet is finally labeled as either benign or malware, i.e., we consider a binary classification problem. The total dataset consists of one million samples (i.e., trimmed packets), half of them benign and half of them coming from the malware traces. We split the dataset on 80\% of the samples for training, 10\% for validation and 10\% for testing purposes. We built the model using the Keras framework running on top of TensorFlow. 

Fig. \ref{roc_results} presents the initial results obtained by the Raw Packet DL model in the detection of malware packets, in the form of a ROC curve. The model is compared to a Random Forest (RF) model, using exactly the same input features and an internal architecture of 100 trees. We choose a RF model based on the generally outstanding detection performance shown by the model in previous work \cite{Casas:2016:PSM:2976749.2989069}, using domain expert input features. DeepMAL can detect more than 70\% of the malware packets with a false alarm rate below 3\%, outperforming the RF model. Indeed, when applying the RF model with the same set of raw input features, the obtained results are very poor. These results point to the ability of the DL-based model to better capture the underlying statistics of the malware, without requiring any specific handcrafted feature set. Still, the absolute detection performance is not good enough to rely on such a DL model with raw packet inputs for malware detection in the practice.

\subsection{Packet vs. Flow Representation Performance}\label{packetvsflow}

We take a step further an consider a similar comparison as before, but considering now both raw packet and raw flow representations as input. We take again publicly available datasets from CTU, but to test on different scenarios, we extend the mix of data by considering multiple \texttt{pcap} files. We consider the same 10 different types of malware captures as before, adding now 16 types of normal captures, choosing the same number of packets for each capture to build a balanced dataset (50\% malware packets and 50\% normal packets).

\begin{table}[t!]
\centering
\footnotesize
{\def\arraystretch{1.8}\tabcolsep=2pt
\begin{tabular}{|c|c|c|c|}
\hline
\textbf{Representation} & \textbf{Dataset size} &$\mathbf{n}$  \textbf{(bytes)} & $\mathbf{m}$ \textbf{(packets)} \\\hline\hline
Raw Packets & $248,850$ & 1024 & -- \\ \hline
Raw Flows & $67,494$ & 100 & 2 \\ \hline
\end{tabular}}
\caption{Parametrization of DeepMAL's input representation.}
\label{tab:parameters}
\end{table}

In Table \ref{tab:parameters} we show the parameters selection for each input representation. All of these parameters where selected again from a statistical analysis of the datasets. Different from the \textit{Raw Packet} representation, we set $n = 100$ bytes for the \textit{Raw Flows} scenario, and keep only the first 2 packets of each flow, i.e., $m = 2$. Note that such a small number of packets has been already proved to be efficient in discriminating among different types of applications when it comes to network traffic classification \cite{class_fly}.

We built two different datasets to fit each one of the considered input representations. The dataset for the \textit{Raw Packet} representation consists of roughly $250,000$ instances. For the \textit{Raw Flows} representation, the dataset consists of about $68,000$ instances. Both datasets are split according to the same scheme as before: 80\% of the samples for training, 10\% for validation and 10\% for testing.

The learning processes for both approaches is described in Fig.~\ref{fig:learning}. In both cases we used mini-batches for the parameters update and we trained the models over several epochs (being an epoch a single pass-through over the complete training dataset). In the case of the \textit{Raw Packets} representation, the training was held over 100 epochs, while in the case of \textit{Raw Flows} we used 10 epochs. We used ADAM as the optimizer function, annealing the learning rate over time. The performance metric chosen in both cases was the accuracy, since the dataset is balanced. For the \textit{Raw Packets} representation we achieved $77.6\%$ of accuracy over the test set; while in the case of \textit{Raw Flows} we achieved an accuracy of $98.6\%$ also over the test set. Note that the learning process performs better when operating at the flow level, as there is some potential over-fitting for this scenario using the packet-level representation. 

Fig.~\ref{fig:rocs} compares the detection performance of both models against a RF model using exactly the same (raw) input features - in the \textit{Raw Flows} case, we flatten the data to fit the input to the RF. In both cases, the internal architecture of the RF consisted of 100 trees using different pruning techniques to prevent over-fitting (e.g., maximum depth, maximum number of instances per leaf, etc.). Once again, we observe a clear out-performance of DeepMAL as compared to the RF models, particularly when operating at the flow level. For \textit{Raw Packets} representations, DeepMAL can detect about 65\% of the malware traffic packets with a false alarm rate below 3\%, with an overall out-performance of nearly 15\% as compared to the RF. Note that in this evaluation, the differences in terms of performance are not as important as in Fig.~\ref{roc_results}. The main differences occur when operating with \textit{Raw Flows} representations, where DeepMAL can detect all malware flows with a false alarm rate below 2.5\%, and more than 90\% of them with less than 0.5\% of alarms. This suggests that, when operating at the flow level, such raw input representation and associated DL architecture can actually provide highly accurate results, applicable in the practice.

For additional benchmarking purposes, we compare the performance of the \textit{Raw Flows} model against a broader set of shallow ML models, using in all cases the computed raw flows inputs for training and testing purposes. These additional models include decision-trees (CART), na\"ive bayes (NB) models, neural networks (MLP), k-nearest-neighbors (k-NN), and support vector machines (SVM). In all cases, we resort to grid-based search to calibrate model parameters. 

Fig.~\ref{fig:rocs_all_cnn} depicts the obtained ROC curves for the tested models. The corresponding AUC values are 0.998 (Raw Flows), 0.928 (RF), 0.944 (SVM), 0.840 (k-NN), 0.835 (NB), 0.930 (MLP), and 0.869 (CART). Results reveal a clear out-performance of DeepMAL as compared now to all the shallow models; indeed, shallow models using raw, non processed input features result in very poor performance.

\begin{figure}[t!]
\centering
\includegraphics[width=0.9\columnwidth]{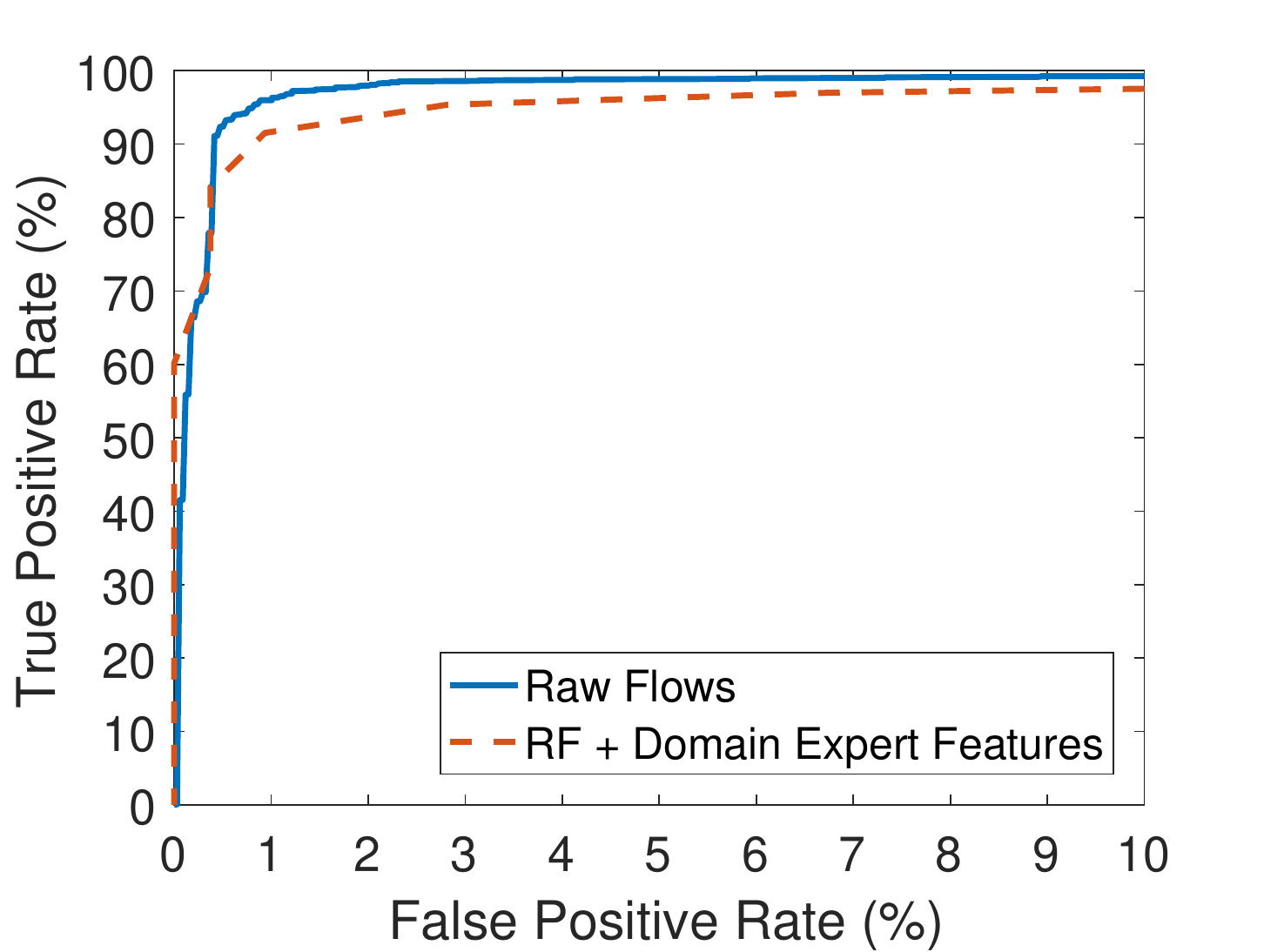}
\caption{\textit{Raw Flows} vs. expert-knowledge-based inputs.}
\label{fig:rocs_expert}
\end{figure}

\subsection{Domain Knowledge vs. Raw Inputs}

The last step of the evaluations tries to answer the third question regarding the goodness and advantages of the proposed approach w.r.t. the standard approach for machine-learning based malware detection. In particular, we study how good is the raw-flows, DL-based model as compared to a RF-based model, the latter using as input specific hand-crafted features based on domain expert knowledge.

The standard approach for detection of malware and network attacks in network traffic is to rely on flow-level features, using traditional in-flow packet measurements such as traffic throughput, packet sizes, inter-arrival times, frequency of IP addresses and ports, transport protocols and shares of specific flags (e.g., SYN packets), etc. We therefore build a set of almost 200 of these features to feed a RF model. Note that besides using traditional features such as min/avg/max values of some of the input measurements, we also consider their empirical distribution, sampling the empirical distribution at many different percentiles. This provides as input much richer information, as the complete distribution is taken into account. We take the same dataset used in Sec. \ref{packetvsflow} for training and testing purposes.

Fig.~\ref{fig:rocs_expert} reports the results obtained in this scenario. Not surprisingly, the RF using expert domain features achieves highly accurate detection performance, detecting about 97\% of all the malware flows with less than 1\% of false alarms. However, also in this scenario, DeepMAL slightly outperforms this domain expert knowledge based detector. As such, we can conclude that the proposed DL architecture can perform as good as a more traditional shallow-model based detector for detection of malware flows, without requiring any sort of expert handcrafted inputs. This of course, shows the great contribution of DeepMAL.

Based on the three sets of evaluations, we can conclude that the proposed DL model, in particular when using raw flow representations as input, can: (i) provide highly accurate and applicable-in-the-practice malware detection results, (ii) capture the underlying malware and normal traffic models better than traditionally used, shallow-like models, and (iii) provide results as good as those obtained through a domain expert knowledge-based detector, without requiring any sort of hand-crafted features.

\begin{table}[t!]
\centering
\footnotesize
{\def\arraystretch{2}\tabcolsep=8pt
\begin{tabular}{|ccc|}
\hline
\textbf{Botnet} & \textbf{Protocol} & \textbf{Activity}\\ \hline\hline
Neris & IRC & spam, click fraud \\ \hline
Rbot & IRC & DDoS \\ \hline
Virut & HTTP & spam, port scan\\\hline
\end{tabular}
}
\caption{Protocols and attacks performed by different kinds of selected botnets in the CTU dataset. Note that both Neris and Virut are spam-based attacks.}
\label{tab:botnet}
\end{table}

\section{From Malware Detection to Classification}\label{sec:multi}

To complement previous malware detection results, in this section we present a variation of the binary classification problem, considering now different sorts of malware traffic as different classes, together with a ``normal'' class representing benign traffic. In the CTU dataset, each capture represents a different \textit{scenario}, in which different sorts of malware were executed using several protocols (e.g., IRC, HTTP, P2P, etc.) and porting different types of attacks (e.g., DDoS, port scan, click fraud, spam, etc.).

To build the dataset we considered three different malware traffic classes, corresponding to three different types of botnets, named Neris, Rbot, and Virut. Thus, our multi-class classification problem has four different classes: three representing malware attacks and one that represents normal activity. The activity and protocols used in the scenarios selected to build our dataset are reported in Table \ref{tab:botnet}. The dataset consists of $160,000$ samples, built in a stratified way, i.e., the dataset is balanced with $40,000$ samples per class.

\begin{table}[t!]
\centering
\footnotesize
{\def\arraystretch{2}\tabcolsep=8pt
\begin{tabular}{|ccccc|}
\hline
\textbf{Class} & \textbf{Accuracy} & \textbf{Precision} & \textbf{Recall} & $\mathbf{F_1}$ \textbf{score}\\ \hline\hline
Normal & $0.878$ & $0.621$ & $0.878$ & $0.727$  \\ \hline
Neris & $0.635$ & $0.814$ & $0.635$ & $0.714$ \\ \hline
Rbot & $0.999$ & $1.000$ & $0.999$ & $1.000$ \\ \hline
Virut & $0.547$ & $0.679$ & $0.547$ & $0.606$\\\hline
\end{tabular}
}
\caption{DeepMAL malware classification performance.}
\label{tab:multimetrics}
\end{table}

For this classification task, we rely on the \textit{Raw Packets} representation. We do so to avoid reducing the dataset, as we need more samples to avoid over-fitting when dealing with multi-class problems. The DL model used for this task is therefore quite similar to the one used for the \textit{Raw Packets} binary scenario. The difference is that now the activation function used in the last fully connected layer is a Softmax function -- a generalization of the binary logistic regression classifier for multiple classes, instead of the sigmoid used for the binary classification. The corresponding loss function is also different for the multi-class classification problem. In this case we used a categorical cross-entropy, and the learning process was held over 50 epochs. The evolution of the learning process, including loss and accuracy after each epoch, is very similar to the previous raw packets-based model (cf. Fig.~\ref{fig:pktlossiter}), as both operate using a similar input representation.

\begin{figure}[t!]
\renewcommand{\arraystretch}{0.7}
\centering
$\begin{array}{cc}
\includegraphics[width=0.85\columnwidth]{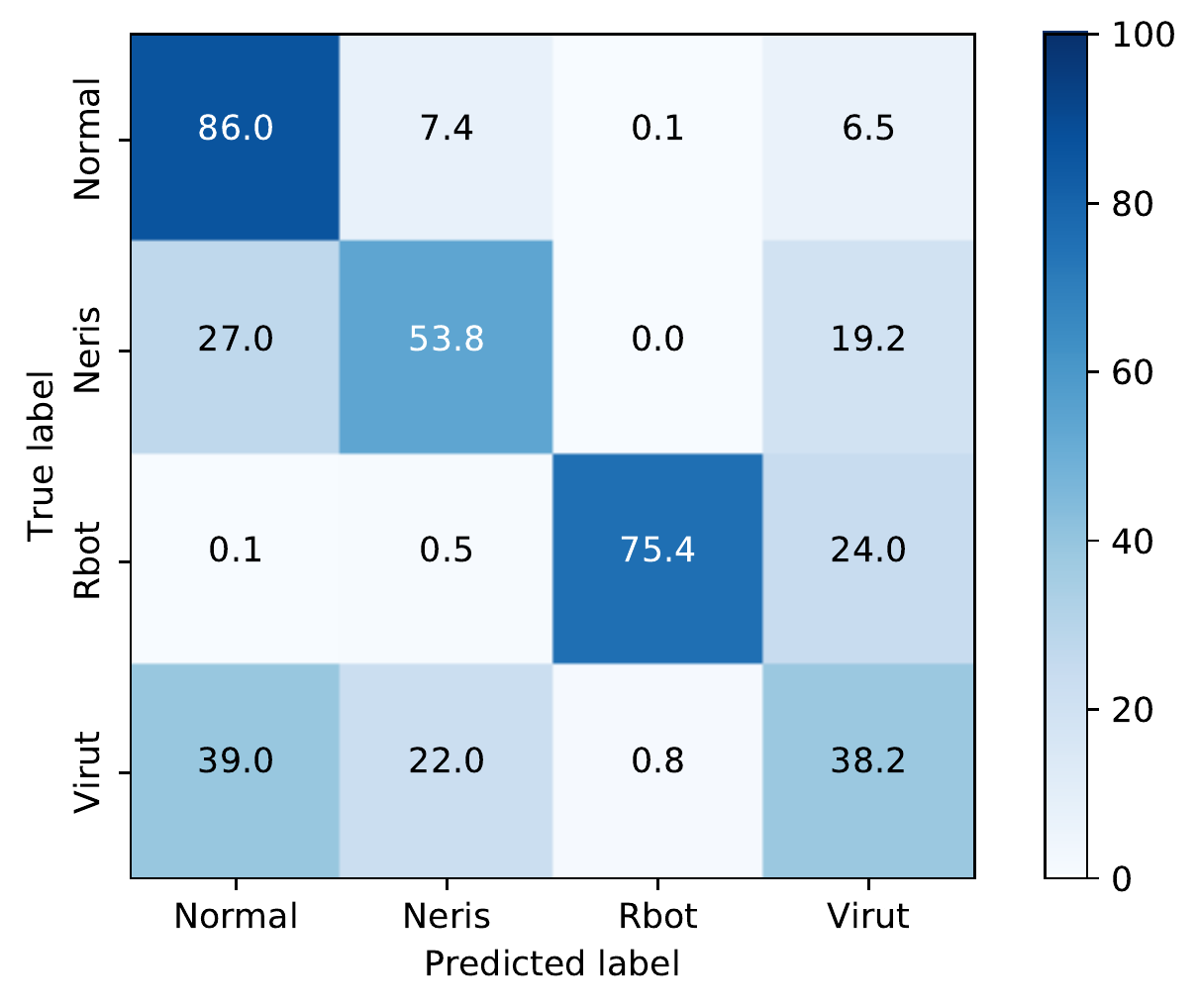}\\
\text{(a) Random Forest.}\\\vspace{-0.2mm}
\includegraphics[width=0.85\columnwidth]{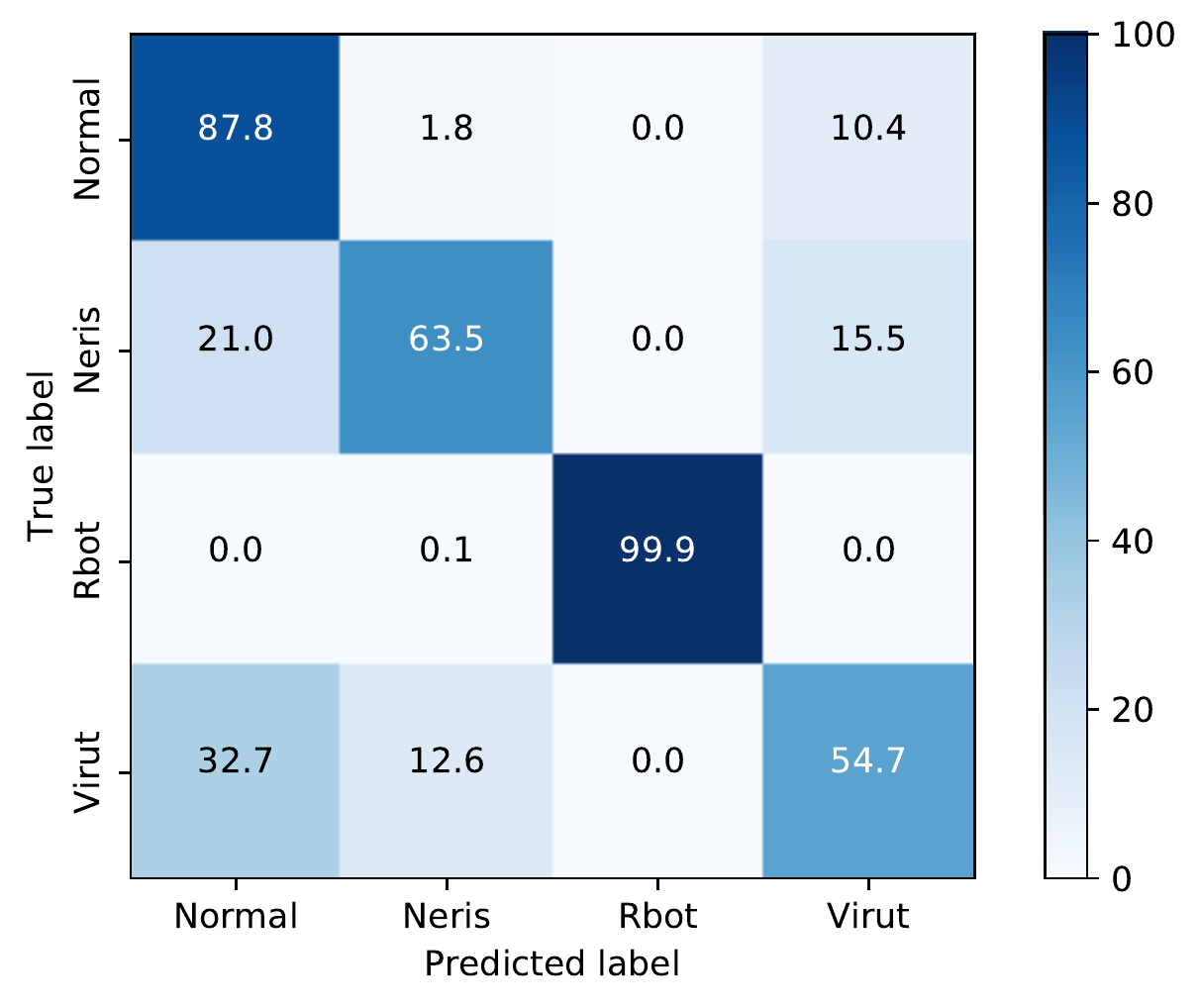}\\
\text{(b) DeepMAL Multi-Class.}\\
\end{array}$
\caption{Normalized confusion matrices (showing percentage values) for both Random Forest and DeepMAL.}
\label{fig:cms}
\end{figure}

\balance

We also compare the performance of DeepMAL against a RF, using the same input features. In Fig.~\ref{fig:cms} we show the normalized confusion matrices for both models -- note that values are shown in percentages; the DL model clearly outperforms the RF for all classes. As a sanity check, note that the accuracy of the multi-class problem considering a binary approach (malware vs. normal) holds similar results as the \textit{Raw Packets} approach presented in Sec.~\ref{packetvsflow} (77.6\% vs. 76.5\%). To complement, in Table \ref{tab:multimetrics} we show additional performance metrics for DeepMAL, including accuracy (AC), precision (PR), recall (RC) and $F_1$ score for each class. Values are computed in a \textit{one-vs-all} schema, in which each class is evaluated against the rest. It is interesting to note that Rbot botnet is detected with an accuracy of 99.9\%, while Neris and Virut achieve 63.5\% and 54.7\% each. It is likely that the fact that both Neris and Virut share spam as an activity attack, could actually be the root-cause behind the bad performance to distinguish one from each other - see Table \ref{tab:botnet}. Interesting is the fact that again, DeepMAL clearly outperforms the RF model, for all variations of malware. As a general conclusion of the multi-class problem, we observe that the raw packets representation does not provide discriminative enough input for the proposed DL architecture to realize properly classification results, but it still outperforms the benchmark RF model. As part of our ongoing efforts, we are working on both testing the multi-class problem using \textit{Raw Flows} representations, as well as on improving classification performance for the raw packets input.

\section{Conclusions}\label{sec:conclu}

In this paper we have presented two different approaches using DL for the detection of malware network traffic, considering \textit{raw} representations of the input network data. Different from traditional, shallow-based approaches, our models operate with raw, byte stream inputs, without requiring any type of handcrafted, expert domain knowledge-based input features or feature engineering, providing an extremely powerful approach. We proposed two different DL architectures, relying on a combination of convolutional and recurrent neural network layers, capable to learn spatio-temporal features out of the raw inputs. Our results show that using \textit{Raw Flows} as input for the DL models achieves much better results than using \textit{Raw Packets}, achieving detection performance which is comparable - or even better, than the obtained by expert-domain knowledge. We also presented a variation of the binary classification model using a multi-class approach to discriminate between different types of malware. In all cases, DL models outperform a strong RF model used as benchmark, as well as other popular shallow-like ML models commonly used in the network security literature, using exactly the same raw input features. This demonstrates the power of the proposed DL architectures and data representations to better capture the underlying statistics of malicious traffic, as compared to more traditional, shallow-like models.


\begin{thebibliography}{99}
\bibitem{DBLP:journals/corr/SimonyanVZ13}
{Simonyan, K., et al.}
\newblock Deep inside convolutional networks: Visualising image classification
  models and saliency maps.
\newblock {\em CoRR abs/1312.6034\/} (2013).

\bibitem{Bengio2013}
{Bengio, Y., et al.}
\newblock {Representation Learning: A Review and New Perspectives}.
\newblock {\em IEEE Trans. P.A.M.I.}, 2013.

\bibitem{Boutaba2018}
{Boutaba, R., et al.}
\newblock {A Comprehensive Survey on ML for Networking:
  Evolution, Applications and Research Opportunities}.
\newblock {\em Journal of Internet Services and Applications}, 9(1):16, 2018.

\bibitem{ml_ad}
{Ahmed, T., et al.}
\newblock ML Approaches to Network Anomaly Detection.
\newblock In {\em Proc. SYSML}, 2007.

\bibitem{survey_adnet}
{Bhuyan, M. H., et al.}
\newblock Network Anomaly Detection: Methods, Systems and Tools.
\newblock In {\em IEEE Comm. Sur. \& Tut.}, vol. 16 (1), pp. 303--336, 2014.
 
\bibitem{survey_ahmed}
{Mohiuddin, A., et al.}
\newblock A Survey of Network Anomaly Detection Techniques.
\newblock In {\em J. of Net. and Comp. App.}, vol. 60, pp. 19--31, 2016.

\bibitem{mlsec_sur} 
{Buczak, A.~L., et al.}
\newblock A Survey of Data Mining and ML Methods for Cyber Security Intrusion Detection.
\newblock In {\em IEEE Communications Surveys \& Tutorials}, vol. 18  (2), pp. 1153--1176, 2008.

\bibitem{MLsurvey}
{Nguyen, T. T., et al.} 
\newblock A Survey of Techniques for Internet Traffic Classification using ML.
\newblock In {\em IEEE Communications Surveys \& Tutorials}, vol. 10  (4), pp. 56-76, 2008.

\bibitem{Casas:2016:PSM:2976749.2989069}
{Casas, P., et al.}
\newblock (Semi)-supervised ML approaches for network security in high-dimensional network data.
\newblock In {\em Proc, ACM CCS}, 2016.

\bibitem{Garcia:2014:ECB:2664689.2665897}
{Garc\'{\i}a, S., et al.}
\newblock An empirical comparison of botnet detection methods.
\newblock {\em Comput. Secur. 45\/}, 2014.

\bibitem{8026581}
{Lopez-Martin, M., et al.}
\newblock Network traffic classifier with convolutional and recurrent neural networks for internet of things.
\newblock {\em IEEE Access 5\/} (2017), 18042--18050.

\bibitem{DBLP:journals/corr/abs-1709-02656}
{Lotfollahi, M., et al.}
\newblock Deep packet: {A} novel approach for encrypted traffic classification using DL.
\newblock {\em CoRR abs/1709.02656\/} (2017).

\bibitem{marin_rawpower_2018}
{Mar\'{\i}n, G., et al.}
\newblock Rawpower: DL based anomaly detection from raw network traffic measurements.
\newblock In {\em Proc. ACM SIGCOMM SRC}, poster, 2018.

\bibitem{marin_wain_2018}
{Mar\'{\i}n, G., et al.}
\newblock DeepSec meets RawPower - DL for Detection of Network Attacks Using Raw Representations.
\newblock In {\em ACM SIGMETRICS Performance Evaluation Review}, vol. 46 (3), pp. 147-150, 2018.

\bibitem{DBLP:journals/corr/abs-1803-10769}
{Radford, B.~J., et al.}
\newblock Network traffic anomaly detection using recurrent neural networks.
\newblock {\em CoRR abs/1803.10769\/}.

\bibitem{class_fly}
{L. Bernaille, et al.}
\newblock Traffic Classification On The Fly.
\newblock ACM CCR, 36(2), pp. 23-26, 2006.

\bibitem{8004872}
{Wang, W., et al.}
\newblock End-to-end encrypted traffic classification with one-dimensional convolution neural networks.
\newblock In {\em Proc. IEEE ISI}, 2017.

\bibitem{flowpic}
{T.~Shapira, et al.}
\newblock FlowPic: Encrypted Internet Traffic Classification is as Easy as Image Recognition  
\newblock In {\em Proc. IEEE INFOCOM Workshops, NI Workshop}, 2019.

\bibitem{7899588}
{Wang, W., et al.}
\newblock Malware traffic classification using convolutional neural network for representation learning.
\newblock In {\em Proc. ICOIN}, 2017.

\bibitem{WangTheAO}
{Wang, Z.}
\newblock The applications of DL on traffic identification.
\newblock In {\em Black Hat USA, Las Vegas\/} (2015).

\bibitem{r1}
{G. Aceto, et al.}
\newblock Mobile Encrypted Traffic Classification Using DL: Experimental Evaluation, Lessons Learned, and Challenges.
\newblock IEEE TNSM, 2019.

\bibitem{r2}
{W. Wang, et al.}
\newblock HAST-IDS: Learning hierarchical spatial-temporal features using deep neural networks to improve intrusion detection.
\newblock IEEE Access 6 (2018): 1792-1806. 

\bibitem{r3}
{G. Aceto, et al.}
\newblock Mobile encrypted traffic classification using DL. 
\newblock TMA Conference 2018.

\bibitem{r4}
{Z. Li, et al.}
\newblock Intrusion detection using convolutional neural networks for representation learning.
\newblock Int. Conf. on Neural Information Processing, 2017.

\bibitem{r6}
{Z. Chen, et al.}
\newblock Seq2img: A sequence-to-image based approach towards IP traffic classification using convolutional neural networks.
\newblock IEEE Big Data, 2017.

\bibitem{r7}
{S. Z. Lin, et al.}
\newblock Character-Level Intrusion Detection Based on CNNs.
\newblock IJCNN, 2018.

\bibitem{r8}
{J. Cui, et al.}
\newblock WEDL-NIDS: Improving Network Intrusion Detection Using Word Embedding-Based DL Method.
\newblock Int. Conf. on Modeling Decisions for AI, 2018.

\bibitem{r9}
{H. Huang, et al.}
\newblock Automatic Multi-task Learning System for Abnormal Network Traffic Detection.
\newblock Int. Jour. of Emerging Tech. in Learning (iJET), pp. 4--20, 2018.

\bibitem{r10}
{O. Salman, et al.}
\newblock A Multi-level Internet Traffic Classifier Using DL.
\newblock 9th NoF, 2018.
\end{thebibliography}
\end{document}